\documentstyle[psfig]{l-aa}
\begin{document}
   \thesaurus{06         
              (02.01.2;  
               08.14.1;  
               13.25.5)} 

\title{The isolated neutron star candidate RBS1223 (1RXS J130848.6+212708)}

   \author{A.D. Schwope\inst{1} \and
           G. Hasinger\inst{1} \and
           R. Schwarz\inst{1} \and
           F. Haberl\inst{2} \and
           M. Schmidt\inst{3}}

   \offprints{A. Schwope}
   
   \institute{Astrophysikalisches Institut Potsdam,
              An der Sternwarte 16, D-14482 Potsdam
\and Max--Planck--Institut f\"ur extraterrestrische Physik,
     Karl--Schwarzschild--Str. 2, 85740 Garching bei M\"unchen, Germany
\and California Institute of Technology, Pasadena, CA 91125, USA}

   \date{Received; accepted}
   
   \maketitle
   \markboth{The INS candidate RBS1223}{Schwope et al.}
   
   \begin{abstract}

In the ROSAT Bright Survey (RBS) we have almost completely optically identified
the brightest $\sim$2000 high-galactic latitude sources from the 
ROSAT All-Sky Survey
Bright Source Catalogue (1RXS). A small number of sources has empty X-ray error
circles on optical images. ROSAT HRI follow-up observations of RBS1223
(=1RXS\,J130848.6+212708), a 
soft object with extreme X-ray to optical flux ratio, have confirmed a 
relatively bright X-ray source, whose position could be determined to an 
accuracy of 1.6 arcsec (90\%) due to the presence of a nearby, X-ray detected 
bright star. Deep Keck R- and B-band images of the field were taken,
but the refined X-ray error circle remains empty to a limiting magnitude 
$B \sim 26^m$. With an X-ray to optical flux ratio of 
$\log (f_{\rm X}/f_{\rm opt})>4.1$ 
this object is almost certainly an isolated neutron star, similar to the 
two so far best-known examples RX\,J1856.4--3754 and RX\,J0720.4--3125. 
We discuss limits on the number of similar objects in the RBS catalogue. 

      \end{abstract}

      \section{Introduction}

The statistics of radio pulsars and supernova explosions as well as the 
abundance of heavy elements suggests that about $10^9$ neutron stars 
must have been born in our Galaxy throughout its life (Narayan \& Ostriker
1990, van den Berg \& Tamman 1991). As young objects, $< 10^4$\,yrs, 
neutron stars appear as rotation-powered bright X-ray and Gamma-ray 
sources or radio pulsars. At intermediate ages, $< 10^6$\,yrs, 
nearby neutron stars can be 
detected by radiation from their cooling surface.
At even later times, $> 10^6$\,yrs, isolated neutron stars (INS) become too 
cold to be detected as a result of internal processes
and could eventually only be seen 
through their photospheric emission in the X-ray band, if they accrete 
enough interstellar matter to heat their surface (Ostriker, Rees \& Silk 
1970). Depending on the assumption about their space velocity distribution,
several thousand old INS could be detectable e.g.~in the ROSAT 
All-Sky-Survey (RASS) (Treves \& Colpi 1991; Blaes \& Madau 1993; 
Madau \& Blaes 1994).

\begin{table*}[t]
\begin{center}
\caption[]{Summary of isolated neutron star candidates}
\begin{tabular}{lcccccccl}
\hline
Name & PSPC & HR1$^1$ & HR2$^2$ & $kT_{\rm bb}$ & $N_H$ & $B$& $\log f{\rm xo}^3$ & Reference\\
     & [cts s$^{-1}$] &  &     & [eV] & [$10^{20}$\,cm$^{-2}$] & [mag] & &  \\
\hline
RX\,J1856 & $3.64\pm0.15$ & $-0.73\pm0.04$ & $-0.92\pm0.07$ & $\sim57$  & 1           & $25.8^4$ & 4.8 & Walter et al.~(1996)\\
RX\,J0720 & $1.69\pm0.07$ & $-0.57\pm0.03$ & $-0.87\pm0.12$ & $\sim79$  & 1           & $>$26  & $>$5.3& Haberl et al.~(1997)\\
RX\,J0806 & $0.38\pm0.03$ & $-0.53\pm0.07$ & $-0.73\pm0.18$ & $78\pm7$  & $2.5\pm0.9$ & $>$24  & $>$3.4& Haberl et al.~(1998)\\
RBS1556   & $0.88\pm0.04$ & $-0.70\pm0.03$ & $-0.58\pm0.10$ & $100\pm10$& $<1$        & $>$22  & $>$3.5& this paper\\  
RBS1223   & $0.29\pm0.02$ & $-0.20\pm0.08$ & $-0.51\pm0.11$ & $118\pm13$& $0.5 - 2.1$ & $>$26  & $>$4.1& this paper\\ 
\hline
\end{tabular}
\end{center}
{\small $^1$ Standard ROSAT hardness ratio HR1: 
($(0.5-2\,{\rm keV})-(0.1-0.4\,{\rm keV})$) / ($(0.5-2\,{\rm keV})+(0.1-0.4\,{\rm keV})$)\\
        $^2$ Standard ROSAT hardness ratio HR2: 
($(0.9-2\,{\rm keV})-(0.5-0.9\,{\rm keV})$) / ($(0.9-2\,{\rm keV})+(0.5-0.9\,{\rm keV})$)\\
        $^3$ X-ray to optical flux ratio $\log(f_{\rm X}/f_{\rm opt})$\\
        $^4$ $V$-magnitude}
\vskip -0.5truecm        
\end{table*}

Recently, two very good INS candidates have been reported: 
RX\,J1856.5--3754 (Walter et al.~1996), RX\,J0720.4--3125 (Haberl et al.~1997). 
Both of them were detected in the RASS as relatively bright sources
(see Tab.~1) with soft, blackbody-like spectra and little interstellar 
absorption. Pointed follow-up observations with the ROSAT PSPC and/or 
HRI yielded very small X-ray error circles which put  
stringent constraints on possible optical counterparts. Walter \& Matthews
(1997) could identify RX\,J1856.5--3754 with a faint blue object in an 
HST observation ($V=25\fm8, U=24\fm5$). The extremely large X-ray to optical 
flux ratio $\log(f_{\rm X}/f_{\rm opt})=4.8$ 
excludes anything other than an isolated 
neutron star for this object. Recent observations of the likely counterpart
of RX\,J0720.4--3125 (star X1) reveals  $B=26\fm1$ (Motch \& Haberl 1998)
and $B=26\fm6$ (Kulkarni \& van Kerkwijk 1998, KvK98). With $B>26^m$
the inferred ratio $\log(f_{\rm X}/f_{\rm opt})>5.3$ 
again indicates an INS origin for the X-ray emission. 
On long time scales the X-ray emission of the two INS candidates is almost 
constant. However, most remarkably, pulsations with a period of 8.4 
seconds were detected for RX\,J0720.4--3125, which, under the assumption
of accretion, indicate a low 
magnetic field ($\sim$10$^{10}$\,G) and a small space velocity 
($\sim$10\,km s$^{-1}$) for the neutron star (Haberl et al.~1997).
A third, fainter candidate INS, RX\,J0806.4--4123, 
also detected in the RASS has 
been reported by Haberl et al.~(1998). Its optical counterpart 
has $B>24^m$ such that $\log(f_{\rm X}/f_{\rm opt})>3.4$. 

Assuming that only two INS with count rates $\sim$2 PSPC cts s$^{-1}$ exist in the RASS,
we expect a number of order 20 over the whole sky with count rates $\sim$0.2\,cts~s$^{-1}$
(c.f.~Neuh\"auser \& Tr\"umper 1998).

 
\section{The ROSAT Bright Survey (RBS)}

In order to improve the determination of the X-ray luminosity function
of X-ray emitting source classes, mainly AGNs, galaxies and clusters 
of galaxies, we have almost completed a program of optical identifications
of bright, high-galactic-latitude  X-ray sources, termed the ROSAT Bright 
Survey RBS. The goal is to completely identify all bright X-ray 
sources, $> 0.2$ PSPC cts s$^{-1}$, 
detected in the ROSAT All-Sky Survey 
(1RXS catalogue, Voges et al.~1996) at galactic latitudes $|b|>30^o$,
excluding LMC, SMC and the Virgo cluster
(Hasinger et al.~1997; Fischer et al.~1998; Schwope et al.~1998).


The RBS is already to 98\% spectroscopically complete.
We find a total of 757 stars, 659 AGN, 297 clusters of galaxies,
171 BL Lac candidates, 43 Galaxies, 49 Cataclysmic variables or X-ray binaries.
Only 36 of the more than 2000 sources in the sample are not yet identified.
Among the known types of X-ray emitters, the BL Lac objects 
are those with the most extreme X-ray to optical flux ratio. 
Yet the optically faintest BL Lac objects at $V \simeq 20\fm5$ have 
$f_{\rm X}/f_{\rm opt} < 10^2$ (Fig.~\ref{f:colcol} shows the 
ratio $f_{\rm X}/f_{\rm opt}$ for BL Lac objects, INS candidates and 
unidentified RBS sources). 

For all unidentified X-ray sources we have produced finding charts from the 
ROE/NRL and APM digitised sky survey catalogues. 
The position errors of bright RASS-sources 
are of sufficient quality that we can immediately localize 
X-ray sources with empty RASS error circles on the sky survey 
plates. For quite a number of such objects we have been awarded short 
follow-up observations with the ROSAT HRI 
in the last years or looked in the archive, in order to 
investigate whether the source is a fluke or has an exceptionally large
X-ray position error (a number of such objects existed in the first 
processing of the all-sky-survey, RASS-I). Another possibility is that the 
X-ray source is extended (cluster or group) and therefore has a formally
empty error circle. Finally, in one case so far the HRI confirmed the position
and flux of the X-ray source and has put very strong constraints on the 
magnitude of an optical counterpart.

\section{X-ray observations of RBS1223} 

\begin{figure}[htp]
\vskip -0.5truecm
\unitlength1cm
\begin{center}
\begin{minipage}[t]{8cm}
\begin{picture}(8,8)
\centerline{\psfig{figure=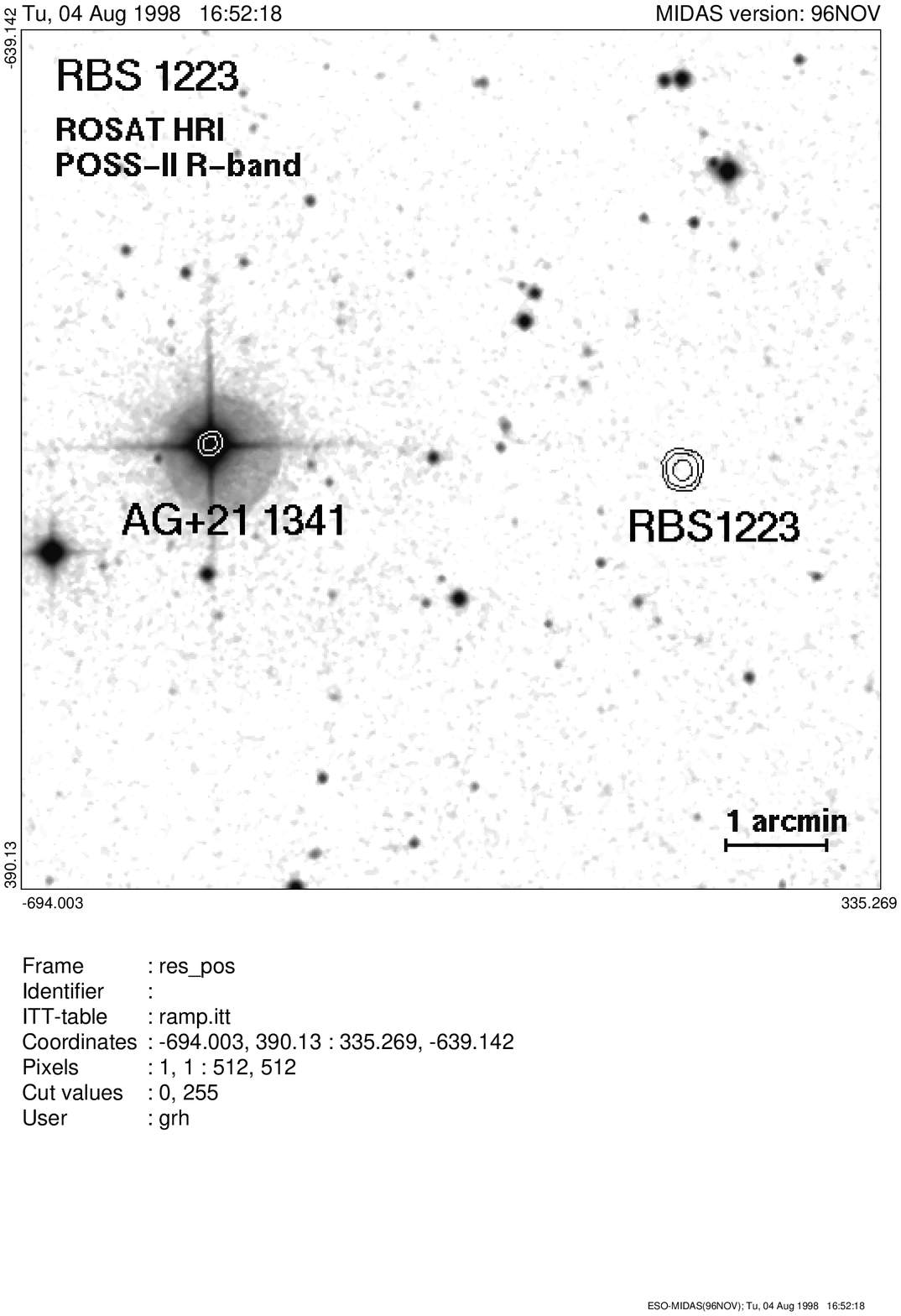,width=8.0cm,clip=}}
\end{picture}\par
\end{minipage}
\end{center}
\vskip -0.5truecm
\caption[ ]{\small ROSAT HRI X-ray contours of the observation of RBS1223 
superposed on the digitised Poss-II R-band image. 
A second HRI source 4.5 arcmin to the East of RBS1223 could fortuitously be 
identified with the 8th magnitude star AG+21 1341}
\label{f:xcont_over}
\vskip -0.5truecm
\end{figure}

The field of RBS1223 has been observed with the ROSAT HRI 
on June 13, 1997 at 19:58 UT for a net exposure time of 2218s. 
We have corrected the individual X-ray photons for the HRI plate-scale 
factor of 0.9972 (Hasinger et al. 1998) and then analysed the data 
using the standard interactive analysis system EXSAS
(Zimmermann et al. 1994). 
The HRI observation showed a point-like X-ray source consistent with the
position of 1RXS\,J130848.6+212708. 
Positions
of ROSAT pointed sources have a routine accuracy of $\sim$10\arcsec\ due to 
inherent attitude errors. The HRI position of RBS1223 however,
could be related to an absolute celestial reference system through the 
fortuitous detection of a second, fainter HRI source about 4.5 arc\-min
west of RBS1223, which could be identified with the bright ($V=8\fm2$) star 
AG+21 1341 (using SIMBAD). The X-ray position of the second HRI source is 
within $0.8\pm0.7\arcsec$ 
of the optical position of AG+21 1341, as determined from 
the APM catalogue. We did not consider it necessary to correct for this shift. 
The final position of RBS1223 is therefore 
RA(2000)$=13^h08^m48.17^s$, 
DEC(2000)$=21^o27'07.5\arcsec$. 
Taking into account the statistical accuracy of the centroiding
for both sources, the 90\% position error for RBS1223 is 1.6 arcsec. 
Fig.~\ref{f:xcont_over} shows a  superposition of the X-ray contours 
of the RBS1223 field on a digitised copy of the 
POSS-II plate, clearly indicating an empty error box down to the plate 
limit. 

The HRI count rate of RBS1223, $0.12\pm0.01$~cts~s$^{-1}$, 
is comparable to the PSPC count rate of $0.29 \pm 0.02$~cts~s$^{-1}$. 
RBS1223 has PSPC hardness ratios which can be fit with a
blackbody temperature of $118\pm13$\,eV, and an interstellar absorption
column density of $N_{\rm H} = (0.5 - 2.1) 
\times 10^{20}$~cm$^{-2}$, consistent with the
total galactic value in this direction, $N_{\rm H, gal} = 2.1
\times 10^{20}$~cm$^{-2}$.
The unabsorbed blackbody flux at earth is about 
$4.5 \times 10^{-12}$~erg~cm$^{-2}$~s$^{-1}$. 
We have analysed the $\sim 266$ HRI
photons from RBS1223 for periodicities in the X-ray flux, but did not 
detect any significant signal in the Fourier transform of the data.


\section{Optical observations of RBS1223}

\begin{figure}[htp]
\vskip -0.5truecm
\unitlength1cm
\begin{center}
\begin{minipage}[t]{8cm}
\begin{picture}(8,8)
\centerline{\psfig{figure=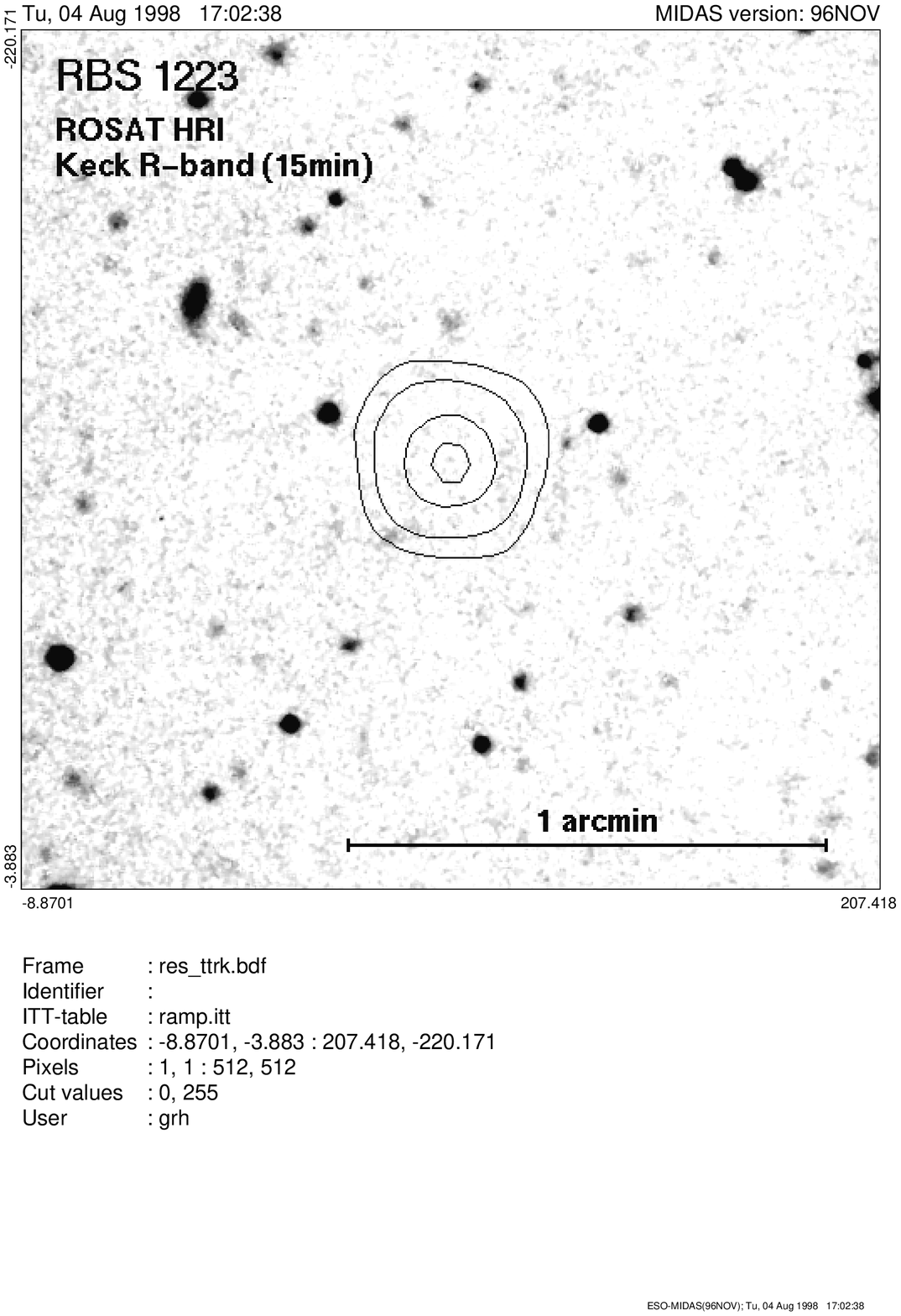,width=8.0cm,clip=}}
\end{picture}\par
\end{minipage}
\end{center}
\vskip -0.5truecm
\caption[ ]{\small ROSAT HRI contours of RBS1223 
superposed on a 15 min R-band Keck II exposure. 
Contours are at 4, 11, 45 and 90\% of the peak emission.
The radius of the innermost 
contour is about 2.5 arcsec, i.e. larger than the actual error circle. No 
optical counterpart is seen to the plate limit of $R \sim 26$}
\label{f:xcont_detail}
\vskip -0.5truecm
\end{figure}

Images of the RBS1223 field have been obtained with the Low Resolution 
Imaging Spectrometer (Oke et al.~1995) at the Cassegrain focus of the
Keck II telescope on the photometric night of March 19th, 1998. 
The $2048 \times 2048$ pixel Loral CCD was read out in single 
amplifier mode. A 15 min exposure through the R filter was taken at
13:33 UT and a 15 min exposure through the B filter at 13:52 UT.  
The reduction was done using MIDAS. The two frames were bias subtracted,
cleaned of cosmic rays and flat-fielded using twilight flats. No 
photometric standards were taken. Using the brightness limits
derived by KvK98 
for the same instrumental setup and same observing
conditions we estimate the 
limiting magnitude of our images at about $26^m$
by scaling with the sqaureroot of the integration time. For our estimate
of the X-ray to optical flux ratio we used the conservative limit 
of $B = 25\fm5$.
Fig.~\ref{f:xcont_detail} 
shows the HRI X-ray contours of RBS1223 superposed
on the $R$ Keck image. No optical counterpart
is detected in the X-ray error circle in either the $R$ or the $B$ image.
Using the observed X-ray flux from above we derive a lower limit
of $\log (f_{\rm X}/f_{\rm opt} > 4.1$, 
using the relation $f_{\rm opt}=10^{-0.4(V+13.42)}$
(Maccacaro et al.~1988) under the assumption of $V=B$.

\section{Discussion}

RBS1223 has all properties that single out the new class of isolated
neutron stars identified by ROSAT: a very high X-ray to optical
flux ratio, a rather soft X-ray spectrum and a constant X-ray 
flux over long time scales. With its very small X-ray error box
it will provide a further challenge for optical identification.
Scaling from the counterpart of RX\,J1856.5--3754 (Walter \& Matthews 1998)
we expect its optical magnitude in the range $R=28^m-29^m$, within reach
of the HST and ground-based 8--10m telescopes. 
The direct prediction of the optical brightness assuming pure blackbody 
radiation with the spectral parameters of Tab.~1 gives only $V = 31\fm5 - 
32\fm5$, beyond the limit of current telescopes.

\begin{figure}[htp]
\vskip -0.5truecm
\unitlength1cm
\begin{center}
\begin{minipage}[t]{8cm}
\begin{picture}(8,8)
\centerline{\psfig{figure=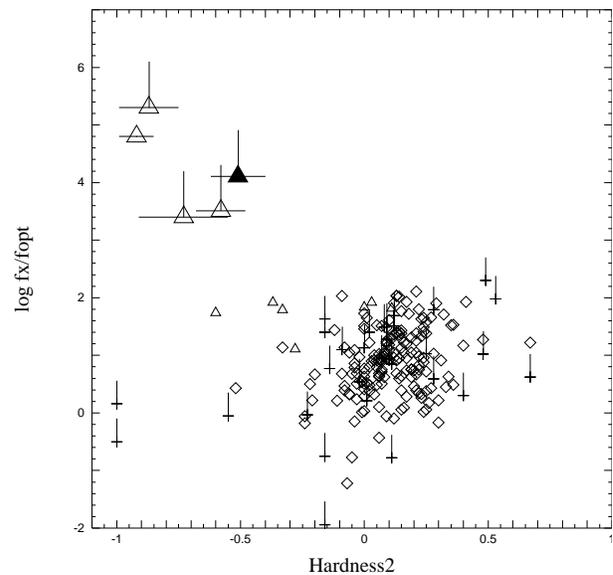,bbllx=61pt,bblly=263pt,bburx=522pt,bbury=702pt,width=8.0cm,clip=}}
\end{picture}\par
\end{minipage}
\end{center}
\vskip -0.5truecm
\caption[ ]{\small X-ray/optical colour colour diagram for selected point-like
RBS sources.  $\log(f_{\rm X}/f_{\rm opt}$) is plotted against the 
hardness ratio HR2, defined in the standard ROSAT way 
(see Tab.~1). Diamonds are BL Lac objects (or candidates) and plus signs are
unidentified objects. For the latter the optical magnitude of the brightest 
object in the error circle, or $V=20\fm5$ for empty error circles has
been assumed. The INS objects (or candidates) listed in Tab.~1
are shown with large triangles (filled symbol for RBS1223).
Other INS candidates from the RBS are shown with small triangles}
\label{f:colcol}
\vskip -0.5truecm
\end{figure}

Fig.~\ref{f:colcol} shows the extreme position of RBS1223 and the other INS
candidates of Tab.~1 plotted in a kind of X-ray/optical colour-colour diagram
together with the otherwise most extreme point-like RBS sources, the BL Lac 
objects. Only sources with 
a very peaked X-ray-optical spectrum can be in the upper left corner of 
this diagram. 
RBS1223 is the faintest of all known INS candidates
and has also the highest X-ray temperature. 
Among the 36 unidentified RBS sources, there is only a handful of 
objects with X-ray hardness ratios consistent with INS. (We have,
however, to be careful about possible misidentified objects in 
X-ray surveys). 
The next best INS candidate from the ROSAT Bright Survey is 
RBS1556, which lies in the same part of the diagram. 
At somewhat lower values of $\log (f_{\rm X}/f_{\rm opt})$ there are a number
of unidentified objects, which could be either BL Lac candidates
or INS. Assuming between 2 and 10 candidate objects in the RBS, 
a rough estimate of the surface density of INS brighter than 0.2
PSPC cts/s is $0.5-1.8$~sources~sr$^{-1}$.
The lower and upper limit can
provide strong constraints on the INS $\log N - \log S$ function and therefore
on the nature of the population as a whole.

The accreting old neutron star scenario was recently criticized  
because of a number of highly uncertain assumptions or unlikely requirements
(see the discussion in KvK98; Wang et al.~1998). 
A very low magnetic field is required to overcome the propeller effect 
and allow accretion. ISM accretion would most likely be highly variable 
on time-scales of months to years. Finally,  
accreting neutron stars would quickly build up a hydrogen-rich envelope,
which should appear much brighter in the optical band than is detected 
for INS (Wang et al.~1998).

Several authors have therefore proposed alternatives or variants for  
the nature of isolated neutron star candidates.
Haberl et al.~(1997) suggest a 
connection of RX\,J0720.4--3125 to the class of anomalous 
pulsars (Mereghetti \& Stella 1995) which might be born
with low space velocities and magnetic fields.
KvK98 discuss the interesting possibility
that the INS candidates are cooling magnetars, very high magnetic
field isolated neutron stars with long spin period, which
are associated with soft gamma ray repeaters.
At any rate, the detection of more objects of this class and
their detailed study with existing and upcoming X-ray and
optical instruments is of vital importance for the 
understanding of their nature. 

\begin{acknowledgements}
We thank the referee Frederick M.~Walter for helpful comments.
The ROSAT project is supported by the Bundesministerium
f\"ur Bildung, Forschung und Wissenschaft (BMBF), by the National
Aeronautics and Space Administration (NASA), and the Science
and Engineering Research Council (SERC). The W.M.~Keck Observatory
is operated as a scientific partnership between the California
Institute of Technology, the University of California, and the
National Aeronautics and Space Administration. It was made possible
by the generous financial support of the W.M.~Keck Foundation.
This research has made use of the SIMBAD database operated at CDS,
Strasbourg, France and was greatly facilitated by use of the APM
catalogue based on scans of the POSS plates performed at the Institute
of Astronomy, Cambridge, UK and the COSMOS scans of the ESO/SERC J 
plates performed at the Royal Observatory Edinburgh.
This work has been supported in part by DLR grant 50 OR 9403 5 (ADS, GH)
and by NASA grant NAG5-1531 (MS). 
\end{acknowledgements}

\end{document}